\documentclass[twocolumn,pra,floatfix]{revtex4}

\usepackage{amsmath}
\usepackage{amstext}
\usepackage{latexsym}
\usepackage{psfig}
\usepackage{graphicx}
\usepackage{amsfonts}
\usepackage{mathrsfs}
\newcommand{\ket}[1]{\left\vert#1\right\rangle}
\newcommand{\bra}[1]{\left\langle#1\right\vert}
\newcommand{\braket}[1]{\left\langle#1\right\rangle}
\begin{document}

  \title{Production of superpositions of coherent states in traveling 
  optical fields with inefficient photon detection}
 \author{H. Jeong, A.P. Lund, and  T.C. Ralph}
  \affiliation{
Center for Quantum Computer Technology, 
Department of Physics, University of Queensland, St Lucia, Qld 4072,
   Australia}

    \date{\today}  

\begin{abstract}
We develop an all-optical scheme to generate superpositions
of macroscopically distinguishable coherent states
in traveling optical fields. It non-deterministically distills
coherent state superpositions (CSSs) with large amplitudes
out of CSSs with small amplitudes using inefficient photon detection.
The small CSSs required to produce CSSs with larger amplitudes
are extremely well approximated by squeezed single photons.
We discuss some remarkable features of this scheme:
it effectively purifies mixed initial states
emitted from inefficient single photon sources
and boosts negativity of Wigner functions of quantum states. 
\end{abstract}

  \maketitle

\section{Introduction}
Schr\"odinger's cat paradox is a famous illustration of the 
principle of superposition  in quantum theory \cite{Schr}.
It poses the question of whether 
a classical object on the macroscopic level can be in a 
state of quantum superposition.
The component states 
composing such a superposition 
should give macroscopically distinct  measurement outcomes \cite{Leggett,Reid}.
A superposition of two optical coherent states
with sufficiently large amplitudes 
of a $\pi$ phase difference 
 is considered a realization of 
such a macroscopic superposition and often called 
a ``Schr\"odinger cat state''. 

Recently, such coherent state superpositions (CSSs)
 in free propagating optical
fields have been found to be useful for various applications
to quantum information processing
\cite{Enk,JKL01,Wang,BaAnTeleport,BaAnW,JK,Ralph,Ralph2,JKpuri,Glancy}.
Quantum teleportation 
\cite{Enk,JKL01,Wang,BaAnTeleport,BaAnW},
quantum computation \cite{JK,Ralph,Ralph2},
entanglement purification \cite{JKpuri} and concentration \cite{JKL01},
error correction \cite{Glancy}, and
remote entangling \cite{BaAnW}, have been extensively studied with CSSs.
In particular, it was shown that quantum computation can
be realized using only linear optics and photon counting, 
given pre-arranged CSSs as resources \cite{Ralph,Ralph2}.
In this approach, a qubit is defined to be a superposition of two coherent
states,
and all four Bell states can be perfectly well discriminated
by photon counting measurements and a beam splitter \cite{JKL01,JKpuri}.
This enables one to construct quantum gates in a relatively 
simple way 
based on the teleportation protocol \cite{Ralph2}.
The amplitudes of coherent states
for qubits and resource CSSs
 should be carefully chosen 
 for efficiency of quantum information processing. 
The CSSs of amplitude  $\alpha >  2$ are required as resources
for efficient quantum computation with simple optical
networks \cite{Ralph2}.

It is known to be extremely hard to generate a 
free propagating CSS using current technology.
It is well known 
 that the CSS can be generated from a
coherent state by a nonlinear interaction in a Kerr medium
\cite{Yurke}. However, Kerr nonlinearity of currently available
nonlinear media is extremely small compared with the level required to generate a CSS
and attenuation in the media is not negligible
 \cite{Boyd}.

Some alternative methods have
been studied to generate a superposition of macroscopically
distinguishable states based upon conditional measurements
\cite{Song,Dakna}. A crucial drawback of these schemes
is that highly efficient photon detection is necessary. 
Both Song {\it et al.}'s scheme \cite{Song} and Dakna {\it et al.}'s  one 
\cite{Dakna} require
photon number resolving measurements which is extremely demanding
 using current technology.
Some other schemes  \cite{Dakna2} 
 require many single photon detectors instead of one $n$-photon
counting detector. 
Even though it is known that many perfect single-photon detectors
enable one to perform nearly perfect $n$-photon counting,
such a scheme would suffer a similar difficulty due to detection inefficiency
of many single photon detectors.
Many perfect detectors used to produce macroscopic superpositions 
can be replaced with two $n$-photon
Fock states and two perfect detectors
\cite{Dakna3}. This
employs another unavailable
factor (two $n$-photon Fock states) by current technology.
A modified scheme \cite{Montina} of
Ref.~\cite{Song} suggested by Montina
can be robust to detection inefficiency under certain conditions 
where success probability is extremely low and
amplitudes of the generated CSSs are small such as $\alpha<1$. 
 None of the above schemes based on
conditional measurements are currently feasible
to generate CSSs with high fidelity,
the main difficulty being the unavoidable 
inefficiency of photon detection.

Cavity quantum electrodynamics has been studied to
enhance nonlinear effects to generate macroscopic superpositions \cite{Tu}.
Some success has been reported in creating such superposition states within
high Q cavities in the microwave \cite{MB} and optical \cite{Mon} domains.
 However, most of the schemes suggested for quantum
information processing with coherent states
\cite{Enk,JKL01,JK,Wang,BaAnTeleport,BaAnW,Ralph,Ralph2,JKpuri,Glancy} 
require {\it free propagating} CSSs.

Recently, it was shown that free propagating optical CSSs
with amplitude up to $\alpha=2.5$ and fidelity $F>0.99$ 
can be generated with squeezed single 
photons and simple all-optical operations \cite{Lund04},
where neither efficient photon detection nor $\chi^{(3)}$
nonlinear interactions are required.
It was also found to be resilient to photon production inefficiency to some extent
as its first step effectively purifies initial mixed states
emitted from an inefficient single photon source \cite{Lund04}.
In a more general sense, these examples reveal that the first 
excited energy eigenstates can be converted to 
a superposition of macroscopically distinguishable states
by linear operations and projections.

In this paper, we extensively analyze the scheme in Ref.~\cite{Lund04} and
find that its purification effects can last for further steps.
The non-deterministic CSS amplification scheme
is found to boost non-classicality of quantum states:
even very small amount of negativity can be drastically
increased by this process.
It is also pointed out that
single photon source is not necessary to obtain squeezed single photons
if another non-deterministic technique, photon subtraction \cite{pst} as
demonstrated in a recent experiment \cite{Wenger}, is employed.

This paper is organized as follows.
In Sec.~II, We briefly define and discuss the CSS 
as a macroscopic superposition state with Schr\"odinger's cat paradox.
In Sec.~III, it is shown that a 
CSS with a small coherent amplitude ($\alpha\leq1.2$) and high fidelity
($F>0.99$) can be deterministically generated by squeezing a single
photon. The discussion is motivated by the approach
of Ref.~\cite{Song}. The Wigner functions
of squeezed single photons and CSSs
 are analytically obtained, and they are 
 compared to visualize the effects
 of squeezing on single photons.
 In Sec.~IV, we fully analyze and discuss the CSS amplification 
scheme with beam splitters, auxiliary coherent fields
 and inefficient detectors.
Sec.~V combines the two ideas from Sec.~III and Sec.~IV
to produce CSSs with amplitude $\alpha>2$.
Weak squeezing, beam mixing with an auxiliary coherent field and photon
detecting  with threshold detectors are enough to generate a CSS
with amplitude up to $\alpha=2.5$ and high fidelity ($F>0.99$)
 given a single photon source. 
Pufication effects for an inefficient single photon source
are another remarkable aspect of our scheme, which
will be discussed in Sec.~VI particularly for multiple iterations of the process.
We conclude with some final remarks in Sec.~VII. 
A recent experiment
by Wenger {\it et al.} \cite{Wenger} is briefly addressed from the viewpoint of 
CSS generation. We emphasize that single photon sources are not necessary to generate
CSSs of $\alpha>2$ employing the photon subtraction technique with our amplification scheme.

\section{Superpositions of coherent states 
as macroscopic superpositions
- Can they be called ``Schr\"odinger cats''?}

A CSS can be defined as
\begin{equation}
\label{CSSdefine}
|{\rm CSS}_\varphi(\alpha)\rangle=
N_\varphi(\alpha)(|\alpha\rangle+e^{i\varphi}|-\alpha\rangle),
\end{equation}
where $N_\varphi(\alpha)$ is a normalization factor,
$|\pm\alpha\rangle$ is a coherent state of amplitude $\pm\alpha$,
and $\varphi$ is a real local phase factor. 
The amplitude $\alpha$ is assumed to be real for simplicity without loss of generality.  
In this paper we refer to the magnitude of $\alpha$ as the size of the CSS.  
Note that CSSs  such as 
$|{\rm CSS}_\pm(\alpha)\rangle=N_{\pm}(\alpha)(|\alpha\rangle\pm|-\alpha\rangle)$
are called even and odd CSSs respectively because the even (odd) CSS
always contains an even (odd) number of photons.

In Schr\"odinger's paradox, a classical object (cat) is in a superposition of two 
macroscopically distinguishable states (alive and dead).
Leggett and Garg have shown the incompatibility 
between quantum mechanical prediction and macroscopic realism
 for a macroscopic superposition state
 \cite{Leggett}. 
The same kind of discussions have been made by Reid to show
violation of Bell's inequality when local realism is macroscopically defined \cite{Reid}.
A CSS is in a macroscopic superposition state 
when its amplitude is appropriately large.
It is often referred to as a ``Schr\"odinger cat state''
or simply ``cat state'' albeit there exists some 
dispute over the term.

There are probably two conspicuous characteristics of CSSs 
which may justify the title ``Schr\"odinger cat states''.
Firstly, coherent states 
are known as the most classical states among pure states.
The coherent states were originally suggested
by Schr\"odinger as a quantum analogy of classical particles \cite{Schr2}.
A classical particle can be represented as a point in the phase space while
it is prohibited by the uncertainty principle for a quantum state.
A coherent state provides the most point-like description of a quantum particle
in the phase space among all quantum states. 
Furthermore, the coherent states do not change their localized shapes
 as they move in a harmonic oscillator potential.  
Their Wigner functions are positive-definite and their $P$-function exist
even though they are delta functions \cite{BRbook}.

Secondly, two coherent states are macroscopically distinguishable 
when they are well separated in the phase space.
Homodyne detection can be considered a macroscopic measurement
as it does not resolve individual quanta (photon).
The error probability $P_e$ of discriminating two coherent states
$|\alpha\rangle$ and $|-\alpha\rangle$
by a homodyne detection 
 is \cite{Barnett} 
\begin{equation}
P_e={\rm Erf(\sqrt{2}\alpha)}-\frac{1}{2},
\end{equation}
where ${\rm Erf}(x)$ is the error function.
The error probability $P_e$ corresponds to the probability of
a wrong discernment by the homodyne detection 
due to the overlap between the two coherent states.
The probability $P_e$ is
extremely small as $P_e<3.2\times 10^{-5}$ for $\alpha>2$.
In such a case, a CSS in Eq.~(\ref{CSSdefine}) can be considered 
a superposition between two macroscopically distinguishable states
of a classical system.

The first characteristic explained above 
could be more or less weaker than the second one
as a justification for ``Schr\"odinger cat states''
being an alternative title of CSSs.
The coherent states are, of course, still far from
typical classical objects. 
It was shown that quantum key distribution using
coherent states and homodyne measurements is secure against any
individual eavesdropping attack \cite{coherentQKD}. 
It has also been argued that
a weak measurement of the squared quadrature observable may yield
negative values for coherent states \cite{coherent}.

The virtual cat in Schr\"odinger's paradox is, to be more precise, entangled with 
a microscopic quantum object while a CSS in Eq.~(\ref{CSSdefine})
is in a single mode superposition.
However, an entangled coherent state 
\begin{equation}
\label{ECS}
|{\rm ECS}\rangle\propto
|\alpha\rangle|\beta\rangle+e^{i\varphi}|-\alpha\rangle|-\beta\rangle
\end{equation}
 can be simply generated by dividing 
a CSS using a beam splitter.
Such entanglement in Eq.~(\ref{ECS}) of macroscopically distinguishable states
is perhaps more closely aligned with Schr\"odinger's original concept \cite{Schr}.

We have shown that the error probability of discriminating between two coherent states,
$|\alpha\rangle$ and $|-\alpha\rangle$,
is extremely small for $\alpha>2$, which justifies the CSS
as, at least, a macroscopic superposition state.
This value ($\alpha>2$) is also appropriate for quantum
computation using optical  coherent states \cite{Ralph}.
Therefore, we are particularly 
interested in generating CSSs of $\alpha>2$
in this paper.

\section{Generation of small coherent state superpositions}

There have been some trials to generate macroscopic superpositions
using the optical parametric amplifier and a single photon source
\cite{Martini,Song}.
In this Section, we show how a previous scheme
\cite{Song}
to generate macroscopic superpositions
can be significantly simplified
so that small CSSs
with high fidelity can be deterministically generated
simply by squeezing single photons.

\subsection{Simplified generation of small coherent state superpositions}

The key idea presented in this Section is motivated by the scheme described 
in~\cite{Song}.  This scheme uses a non-linear coupling between two optical modes which 
realizes a quantum
non-demolition (QND) measurement on the $\hat{Y}$ quadrature of one mode.  This is done by 
coupling the mode containing the
signal to another ancillary mode as described in~\cite{LaPorta}.  We define the 
quadrature operators of a single mode in terms of the usual creation and annihilation 
operators as
$
\hat{X} = \hat{a} + \hat{a}^\dagger
$
and
$
\hat{Y} = -i (\hat{a} - \hat{a}^\dagger)
$
so that 
$
[\hat{X}, \hat{Y}] = 2i
$
and hence
\begin{equation}
\label{quad-uncer}
\sqrt{\langle \Delta \hat{X}^2 \rangle \langle \Delta \hat{Y}^2 \rangle} = 1.
\end{equation}
So as to avoid confusion between the two quantized modes of the EM field in the QND appratus, 
the mode which the QND measurement is performed is called the signal mode and the mode it 
interacts with to assist with the measurement is called the meter mode.  Also the operators 
associated with the observables of these modes are labelled with subscript `s' for signal and 
`m' for meter.  The device described in~\cite{LaPorta} uses the two polarization modes of a 
single spatial mode as the signal and meter modes.  For example, the horizontal polarization 
might contain the signal and vertical polarization the meter.  The meter mode is usually 
assumed to be prepared in the vacuum state.  The two polarization modes are mixed by a 
wave-plate by an angle $\theta$.  Then two mode squeezing is performed between the two 
polarization modes by a $\chi^{(2)}$ non-linear crystal. The squeezing parameter $r$ is 
determined by the power applied to a pump beam which creates a squeezed vacuum in the absence 
of any input signal.  Finally the polarizations are mixed by a wave-plate by the same angle 
$\theta$. 
\begin{figure}
\centerline{\scalebox{0.55}{\includegraphics{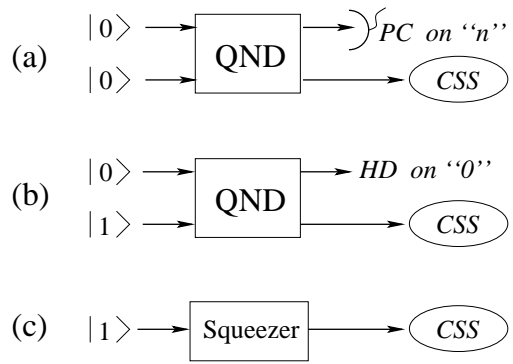}}}
\caption{
A schematic of the simplification of the CSS generation.
$PC$ represents photon counting and $HD$ represents homodyne detection.
(a) Conditional production using QND with photon counting, 
(b) conditional production using QND with homodyne detection,
and (C) deterministic production only by squeezing a single photon.
}
\label{laporta:qnd-dev}
\end{figure}
The evolution through this device is unitary and can be represented by a unitary operator 
$\hat{U}$.  When the squeezing parameter and wave-plate mixing angle are related by $\tanh{r} 
= \sin{2\theta}$ (called the QND condition) then the quadrature operators transform 
as~\cite{Song}
\begin{equation}
\label{qnd_transformation_1}
\left (
\begin{array}{c}
\hat{X}_{s} \\
\hat{X}_{m}
\end{array}
\right)_{o} = \hat{U}^\dagger \left (
\begin{array}{c}
\hat{X}_{s} \\
\hat{X}_{m}
\end{array}
\right)_{i} \hat{U} \left(
\begin{array}{cc}
1 & -2 \sinh r \\
0 & 1 
\end{array}
\right)
\left(
\begin{array}{c}
\hat{X}_{s} \\
\hat{X}_{m}
\end{array}
\right)_{i}
\end{equation}
\begin{equation}
\label{qnd_transformation_2}
\left (
\begin{array}{c}
\hat{Y}_{s} \\
\hat{Y}_{m}
\end{array}
\right)_{o} = \hat{U}^\dagger \left (
\begin{array}{c}
\hat{Y}_{s} \\
\hat{Y}_{m}
\end{array}
\right)_{i} \hat{U} \left(
\begin{array}{cc}
1 & 0 \\
2 \sinh r  & 1 
\end{array}
\right)
\left(
\begin{array}{c}
\hat{Y}_{s} \\
\hat{Y}_{m}
\end{array}
\right)_{i} .
\end{equation}
From these relations it is possible to see that the $\hat{Y}_s$ operator is left unchanged 
through the apparatus but the $\hat{Y}_m$ operator is mixed with the $\hat{Y}_s$ operator.  
This allows information about the $\hat{Y}$ quadrature of the signal to be gathered from the 
meter mode whilst leaving the signal itself undisturbed.  Note that to satisfy the uncertainty 
relation from equation~\ref{quad-uncer} the $\hat{X}$ quadrature of the signal output is not 
identical to its input.  

In attempting to generate CSSs, the scheme in~\cite{Song} suggests preparing the signal (and 
meter) in the vacuum state.  Then apply the QND apparatus just described and perform a photon 
number measurement on the meter.  The signal output is only accepted when a predefined number 
of photons is registered in the measurement.  It is shown heuristically in~\cite{Song} that 
one would expect a CSS to be generated when $r \gg 1$.  When an odd number of photons is 
counted in the meter mode the output is close to an odd CSS and when an even number of photons 
is counted the output is close to an even CSS.  

The scheme described in~\cite{Song} relies on photon counting measurements to post-select the 
desired output state.  This ability to conditionally select the output induces the required 
non-linearity.  However this requires efficient detection and photon number  measurements 
which are difficult to implement.  One possible resolution to this would be to use detection 
schemes which can be made to perform efficiently when post-selecting the output state.  We 
suggest here to use homodyne detection which might be performed with much higher efficiency 
than photon counting measurements.  

Homodyne measurements are effectively a measurement of the $\hat{X}$ or $\hat{Y}$ quadratures 
depending on the phase of a reference signal.  The eigenvalue spectrum of these operators is 
continuous which makes post selection on a particular value have little meaning as that one 
value is infinitesimally small in the set of all possible values.  For example, one could 
measure $\hat{X}$ and select the result for eigenvalue zero but one could never be sure that 
the result was precisely zero.  To circumvent this problem one can use the technique shown 
in~\cite{Branczyk} which accepts events within a given window of possible values and rejects 
all others.

In order to produce states similar to those in~\cite{Song} by homodyne post-selection we 
suggest preparing the signal mode in a Fock state, then post select by performing a homodyne 
measurement on the meter mode after the QND apparatus.  The condition to accept the output of 
the signal is if the measurement was in the range $(-\delta, \delta)$ where $\delta$ is some 
small constant.  
We provide a heuristic description in a similar fashion to that in~\cite{Song} but perform a 
more complete analysis on a simplification that naturally arises when trying to write down the 
output state.

The heuristic description proceeds by expanding the signal mode photon number operator in 
terms of the quadrature operators; i.e.
\begin{equation}
\label{photon_op_signal_in}
\hat{n}_{si} = \frac{\hat{X}_{si}^2}{4}  + \frac{\hat{Y}_{si}^2}{4} - \frac{\hat{I}_{si}}{2}
\end{equation}
where the subscript $s$ represents the fact that the operator represents the signal mode and 
the subscript $i$ for input.  The QND apparatus leaves the $\hat{Y}$ quadrature of the signal 
unchanged by equation~(\ref{qnd_transformation_2}) so
\begin{equation}
\hat{n}_{si} = \frac{\hat{X}_{si}^2}{4}  + \frac{\hat{Y}_{so}^2}{4} - \frac{\hat{I}_{si}}{2}. 
\end{equation}
Here the subscript $o$ represents the operator for the output mode.  Now the $\hat{X}$ 
quadrature of the signal transforms as
\begin{equation}
\hat{X}_{so} = \hat{X}_{si} - 2 \sinh r \hat{X}_{mi}
\end{equation}
and the $\hat{Y}$ quadrature of the meter transforms as
\begin{equation}
\hat{Y}_{mo} = \hat{Y}_{mi} + 2 \sinh r \hat{Y}_{si}
\end{equation}
where the subscript $m$ represents the meter mode.  We now require a post-selective
measurement on $\hat{Y}_{mo}$.  
We use a semi-classical approach to complete this heuristic description by
converting operators back into classical variables. In terms of the semi-classical
quadrature variables, the effect of the post-selective measurement can be included
by setting $Y_{mo} = 0$.  So we can write the signal output as
\begin{equation}
Y_{so} = Y_{si} = -\frac{1}{2 \sinh r} Y_{mi}.
\end{equation}
From this equation we can see that after post-selection the $Y$
quadrature of the signal output is a scaled form of the $Y$ quadrature
to the meter input.  Substituting this expression into
equation \ref{photon_op_signal_in} and rearranging for the signal output
$X$ quadrature one obtains
\begin{equation}
X_{so} = \pm \sqrt{ 4 n_{si} + 2 - \left(\frac{Y_{mi}}{2 \sinh r} \right)^2 } - 2 \sinh r 
X_{mi}.
\end{equation}
Here we consider $n_{si}$ the semi-classical form of the signal input photon number operator 
which will only take on integer values.
The meter input is a vacuum state and hence the $X$ and $Y$ quadratures contain Gaussian 
noise.  The signal input contains a definite photon number.  So the first term under the 
square root will act like a constant.  The second term will have some scaled Gaussian random 
noise and so will the term outside the square root.  However, the multi-valued nature of the 
square root gives the two peaked behaviour required.  This completes our heuristic 
description.

In this paper we will not perform an in depth analysis of this device in full to rigorously 
confirm the results of the heuristic argument just given.   However a complete analysis has 
been performed which confirms this description~\cite{APLThesis}.
Here we will analyze a simplification of this device which has similar functionality and will 
prove fruitful towards achieving our goal of a simplified experiment.  

One can show that the unitary operator which generates the operator transformation 
equations~\ref{qnd_transformation_1} and~\ref{qnd_transformation_2} of the QND apparatus is
\begin{equation}
\label{qnd_unitary}
\hat{U} = e^{i 2 \sinh (r) \hat{X}_m \hat{Y}_s}.
\end{equation}
Now consider this operator acting on the initial state where the signal is in a Fock state and 
the meter is in the vacuum state; i.e.
\begin{equation}
\hat{U} \ket{n}_s \ket{0}_m.
\end{equation}
Inserting two instances of the identity, one expanded over the eigenstates of $\hat{X}_m$ and 
the other eigenstates of $\hat{Y}_s$ so we have
\begin{equation}
\int_{-\infty}^\infty \mathrm{d}x_m \int_{-\infty}^\infty \mathrm{d} y_s \ e^{i 2 \sinh (r) 
\hat{X}_m \hat{Y}_s} \ket{y_s} \ket{x_m} \braket{y_s | n} \braket{x_m | 0}.
\end{equation}
When expressions for the inner products are substituted, this equation becomes
\begin{widetext}
\begin{equation}
\mathscr{N}_n N_\sigma \int_{-\infty}^\infty \mathrm{d}x_m \int_{-\infty}^\infty \mathrm{d} 
y_s \  e^{i 2 \sinh (r) x_m y_s} e^{-y_s^2/2} H_n(y_s) e^{-x_m^2/{2 \sigma}} \ket{y_s}  
\ket{x_m}.
\end{equation}
\end{widetext}
where $\mathscr{N}_n$ and $N_\sigma$ are normalization factors for the $n$th photon term and a 
Gaussian with standard deviation $\sigma$ respectively and $H_n(y_s)$ is the $n$th Hermite 
polynomial.  The process of projecting onto the $\hat{Y}_m = 0$ state can be included by 
taking the inner product of this state with the state on the meter mode alone $\mathrm{d}Y 
\bra{y_m = 0}$.  This leaves us with a $\braket{y_m=0 | x_m}$ inside the integral. This term 
is $e^{i 0 \cdot x_m} = 1$.  Hence the post-selected output state is
\begin{widetext}
\begin{equation}
\mathrm{d}Y \mathscr{N}_n N_\sigma \int_{-\infty}^\infty \mathrm{d}x_m \int_{-\infty}^\infty 
\mathrm{d} y_s \  e^{i 2 \sinh (r) x_m y_s} e^{-y_s^2/2} H_n(y_s) e^{-x_m^2/{2 \sigma}} 
\ket{y_s}.
\end{equation}
\end{widetext}
When the terms involving $x_m$ are grouped together and factored by completing the square and 
the integration over $x_m$ is performed one is left with
\begin{equation}
\mathrm{d}Y \mathscr{N}_n \int_{-\infty}^\infty \mathrm{d} y_s \  e^{-\frac{(1+4 \sigma 
\sinh^2(r))y_s^2}{2}} H_n(y_s)  \ket{y_s}.
\end{equation}
This can be a written a little clearer if we set $\kappa = \sqrt{1+4\sigma \sinh^2(r)}$.  
Hence the signal output is
\begin{equation}
\mathrm{d}Y \mathscr{N}_n \int_{-\infty}^\infty \mathrm{d} y_s \  e^{-\frac{(\kappa 
y_s)^2}{2}} H_n(y_s)  \ket{y_s}.
\end{equation}
This state is not normalised as the process which we have chosen to generate it is 
non-deterministic.  For the special case of $n = 1$ the normalised state is
\begin{equation}
\sqrt{\frac{\kappa^3}{2\sqrt{\pi}}}\int_{-\infty}^\infty \mathrm{d} y_s \  e^{-\frac{(\kappa 
y_s)^2}{2}} H_1(y_s) \ket{y_s}
\end{equation}
and as the Hermite polynomial $H_1(y_s)$ is linear, a $\kappa$ can be 
moved from under the square root to inside the integral to give
\begin{equation}
\sqrt{\frac{\kappa}{2\sqrt{\pi}}}\int_{-\infty}^\infty \mathrm{d} y_s \  e^{-\frac{(\kappa 
y_s)^2}{2}} H_1(\kappa y_s) \ket{y_s}
\end{equation}
which we state is just a re-scaling (or squeezing) of the momentum wave-function of a single 
photon.  So the case of a single photon input into the QND device with the projective 
measurement described above is equivalent to squeezing a single photon Fock state.

\subsection{Squeezed single photons and ideal CSSs}

\begin{figure}
\centerline{\scalebox{0.57}{\includegraphics{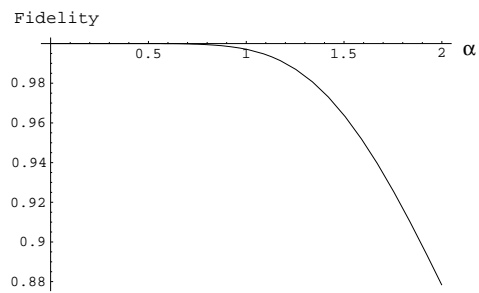}}}
\caption{The fidelity between an odd CSS and squeezed single photon.
The odd CSS is extremely well approximated by the
squeezed single photon for a small coherent amplitude, $\alpha\leq1.2$. }
\label{fid-nopa}
\end{figure}

We have thus been guided to comparing a single mode squeezed single photon state with an odd 
CSS.  The single mode squeezing operator is
\begin{equation}
\hat{S}(r) = e^{-\frac{r}{2}(\hat{a}^2 - \hat{a}^{\dagger 2})},
\end{equation}
where $r$ is the squeezing parameter and $\hat{a}$ is the annihilation operator.
This operator reduces quantum noise of a vacuum state in the phase quadrature by a factor of 
$e^{-r}$.  
When the squeezing operator is applied to a single photon the resultant state
can be expanded in terms of photon number states as
\begin{equation}
\hat{S}(r) \ket{1} = \sum_{n=0}^{\infty} \frac{(\tanh r)^n}{(\cosh r)^\frac{3}{2}} 
\frac{\sqrt{(2n+1)!}}{2^n n!} \ket{2n + 1}.
\label{e-fs}
\end{equation}
The state contains only odd photon numbers and has coefficients decaying 
exponentially as $n$ increases in a similar fashion to an odd CSS.  The 
fidelity of this state with an odd CSS is
\begin{eqnarray}
F(r,\alpha) &=& |\langle {\rm CSS}_-(\alpha)|S(r)|1\rangle|^2\nonumber\\
&=&\frac{2\alpha^2\exp[\alpha^2(\tanh r-1)]}{(\cosh r)^3(1-\exp[-2\alpha^2])}.
\end{eqnarray}
If an odd CSS of size $\alpha$ is desired then the fidelity is maximised when $r$ satisfies
\begin{equation}
\cosh r = \sqrt{\frac{1}{2} + \frac{1}{6} \sqrt{9 + 4\alpha^2}}.
\label{max_cond}
\end{equation}
Fig.~\ref{fid-nopa} shows the maximized fidelity on the y-axis plotted
against a range of possible values for $\alpha$ for the desired odd CSS.
Some example values are:
 $F=0.99999$ for amplitude $\alpha=1/2$,
  $F=0.9998$ for $\alpha=1/\sqrt{2}$, 
and  $F=0.997$ for $\alpha=1$, where the maximizing squeezing parameters are 
$r=0.083$, $r=0.164$ and $r=0.313$ respectively.
These values correspond to $V=0.85$, $V=0.72$ and $V=0.53$, 
where $V$ is the variance of the squeezed quadrature variable.
Firstly note that for $\alpha$
very close to zero the fidelity approaches unity. When $\alpha \rightarrow 0$,
$r \rightarrow 0$ and hence the squeezing operator $\hat S(r)$ approaches
the identity transformation. An odd CSS with $\alpha$ very
close to zero has a significant contribution from a single photon 
and very little from higher odd photon numbers.  This is the reason for the high fidelity as 
$\alpha$ tends to zero.  The fidelity remains high for $\alpha$ near zero as one can match the 
three photon contribution to the CSS by the squeezing operator whilst still being able to 
neglect higher order photon number terms.  Eventually as $\alpha$ increases, higher photon 
numbers cannot be matched and so as $\alpha$ tends to infinity, the fidelity tends to zero.

\begin{figure}
\centerline{(a)}
\centerline{\scalebox{0.42}{\includegraphics{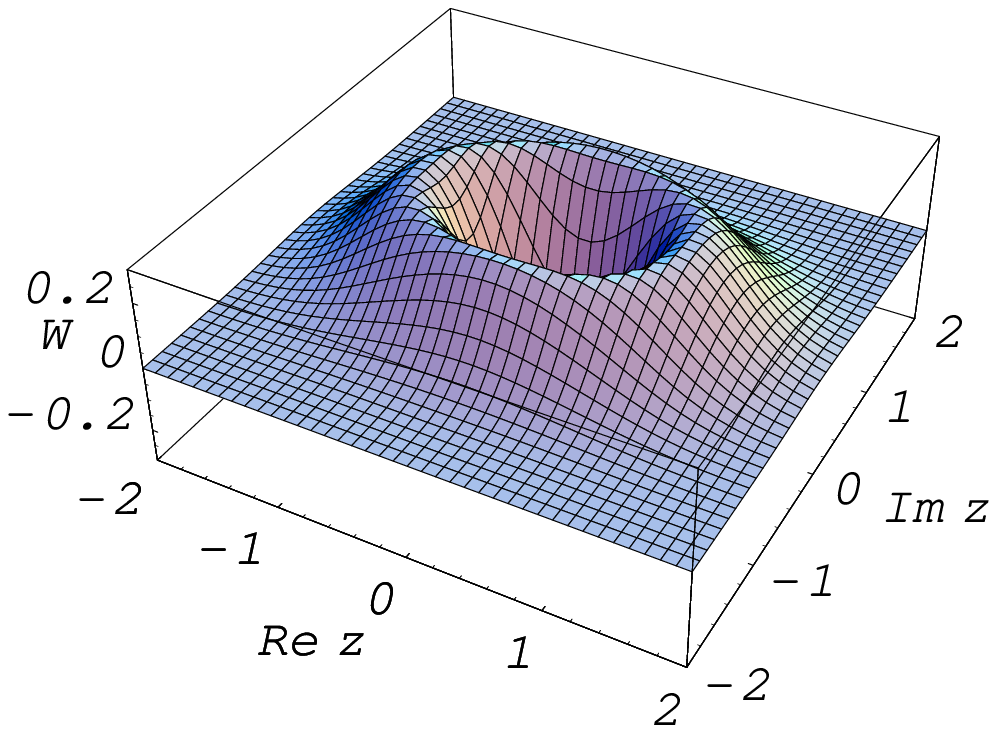}}
\scalebox{0.42}{\includegraphics{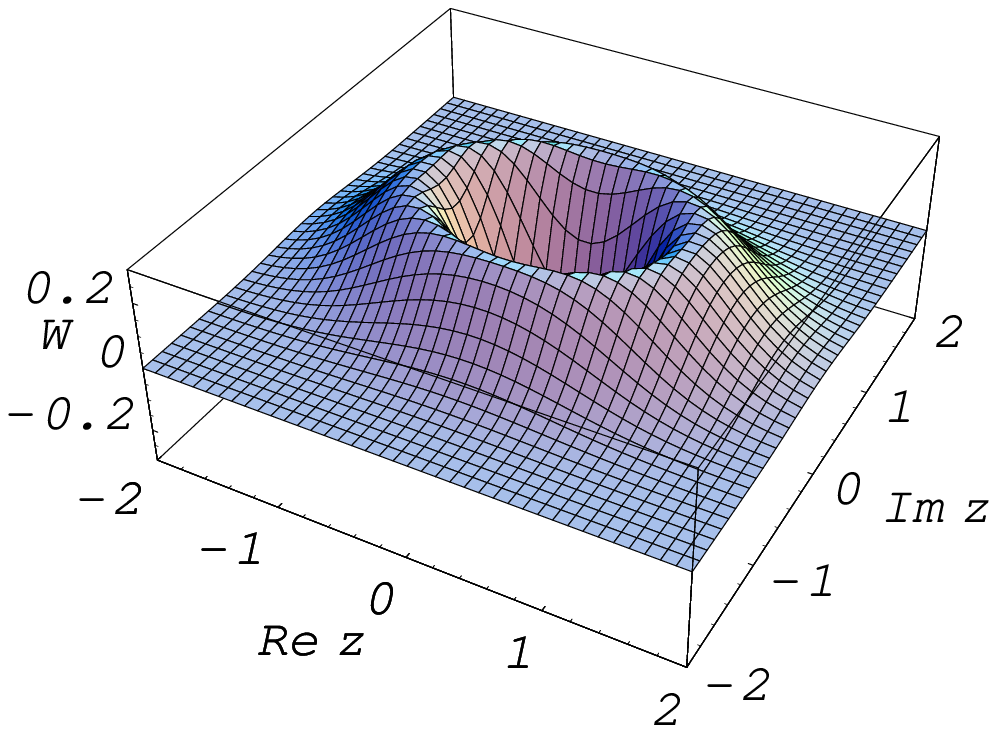}}}
\vspace{0.2cm}
\centerline{(b)}
\centerline{\scalebox{0.42}{\includegraphics{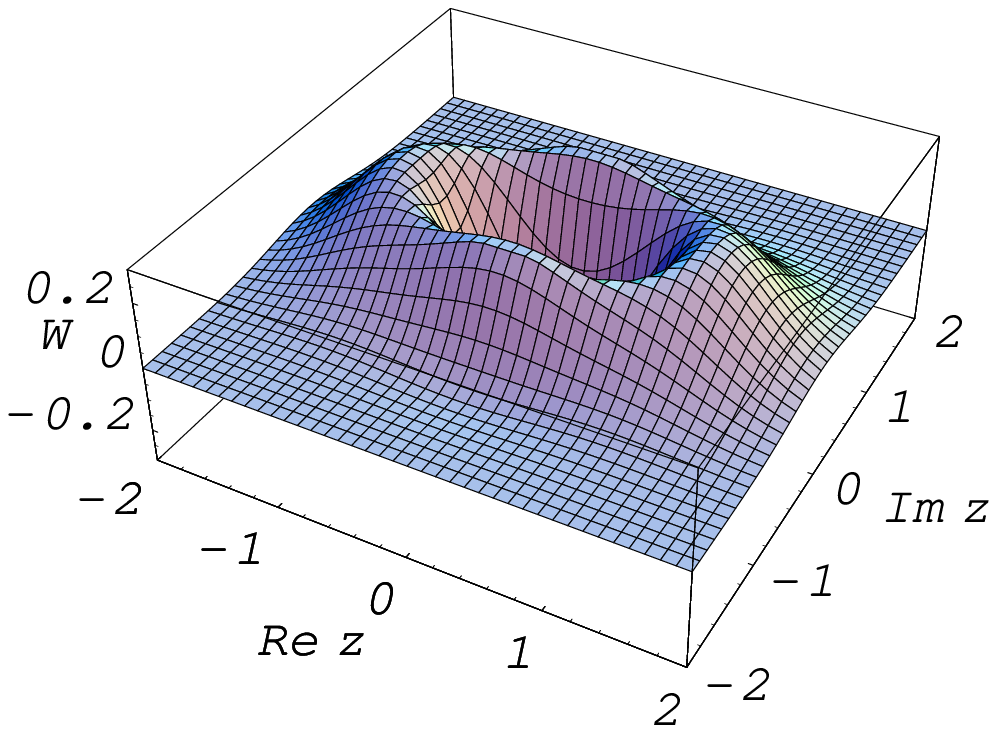}}
\scalebox{0.42}{\includegraphics{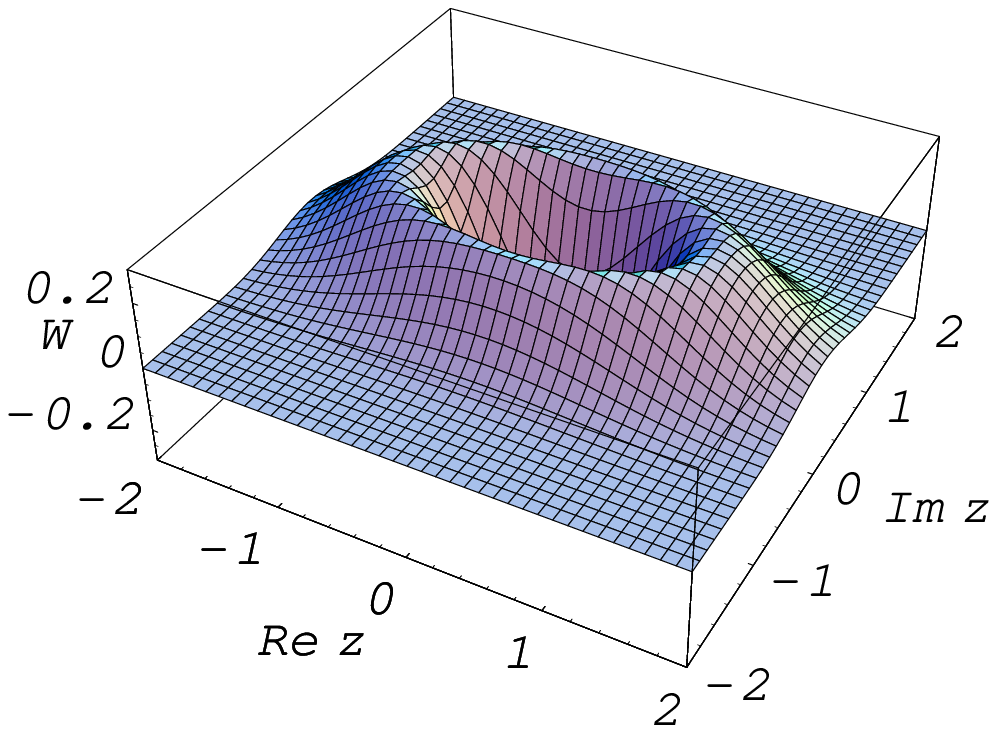}}}
\vspace{0.2cm}
\centerline{(c)}
\centerline{\scalebox{0.42}{\includegraphics{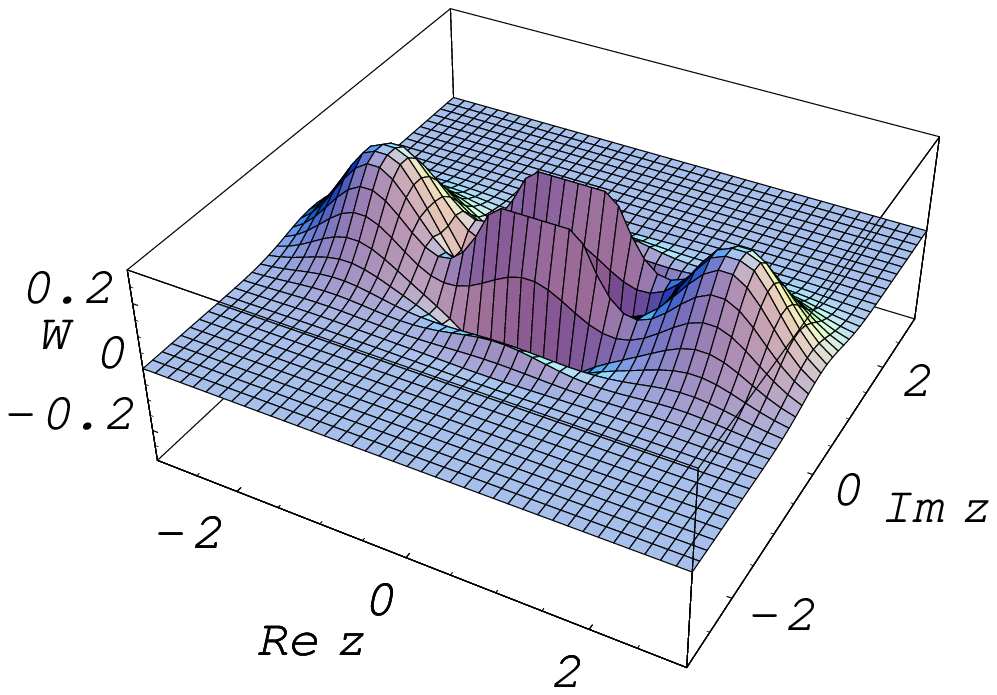}}
\scalebox{0.42}{\includegraphics{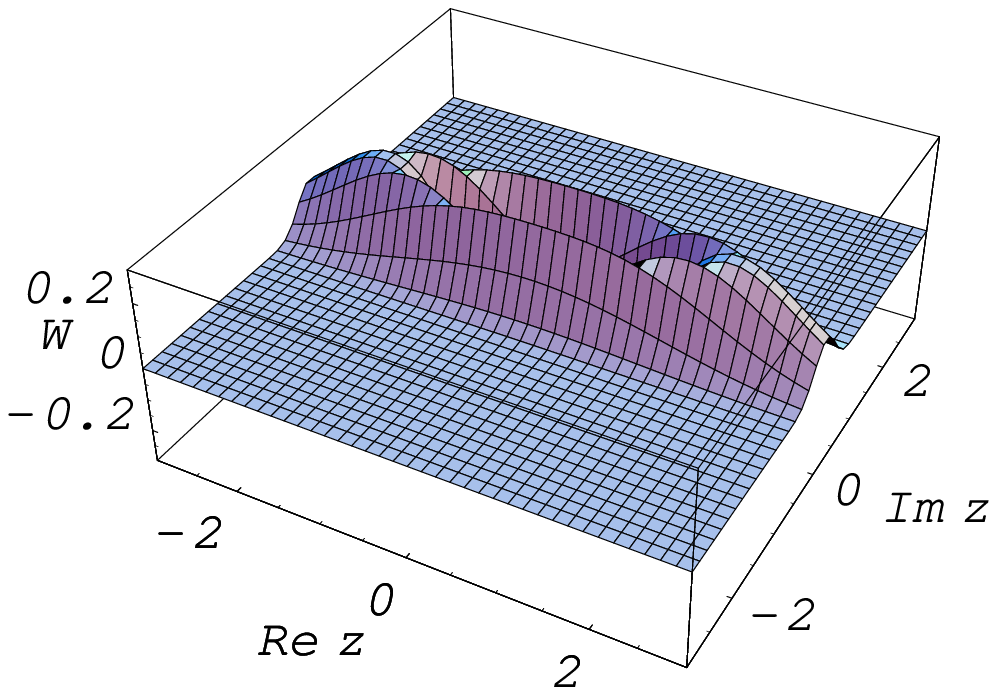}}}
\caption{(Color online)
The Wigner functions of odd CSSs (left) and squeezed single photons (right).
The amplitudes of CSSs are (a) $1/\sqrt{2}$, (b) $1$ and (c) 2.
The degrees of squeezing of squeezed single photons are (a) 0.164, (b) 0.313 and (c) 0.853,
which are chosen for maximum fidelity with CSSs.
It is evident from the figure 
that only small amount of squeezing makes a single photon a good approximate CSS.}
\label{fig-wig}
\end{figure}

The role of squeezing on single photons becomes clear by comparing Wigner functions
of the squeezed single photons and CSSs.
The Wigner function of a squeezed single photon can be obtained
from its characteristic function 
\begin{eqnarray}
&&\chi_s(\eta)={\rm Tr}\Big[S(r)|1\rangle\langle 1|
S^\dagger(r)e^{\eta {\hat a}^\dagger-\eta^*{\hat a}}\Big]\nonumber\\
&&~~~=\exp\big[-\frac{1}{2}(e^{2r}\eta_r^2+e^{-2r}\eta_i^2)
\big](1-e^{2r}\eta_r^2+e^{-2r}\eta_i^2).\nonumber\\
\end{eqnarray}
The Wigner function is then
\begin{eqnarray}
&&W_s(z)=\frac{1}{\pi^2}\int e^{\eta^*z-\eta z^*}\chi_s(\eta)d^2\eta\nonumber\\
&&~=\frac{2}{\pi}\exp[-2(e^{2r }z_r^2+e^{-2r }z_i^2)]
(4e^{2r}z_r^2+4e^{-2r}z_i^2-1).\nonumber\\
\end{eqnarray}
The Wigner function of the CSS is obtained by the same method as 
\begin{equation}
W_c^\pm(z)=\frac{e^{-2|z|^2}}{\pi(1\pm e^{-2\alpha^2})}
\Big\{ e^{-2\alpha^2}(e^{-4\alpha z_r}+e^{4\alpha z_r})
\pm 2\cos 4\alpha z_i \Big\},
\end{equation}
where $W_c^+(z)$ ($W_c^-(z)$) is the Wigner function of the even (odd) CSS.
The Wigner functions of odd CSSs with amplitudes $1/\sqrt{2}$, 1, 2
and the Wigner functions of 
corresponding squeezed single photons are plotted in 
Fig.~\ref{fig-wig}. It shows that 
only small amount of squeezing makes a single photon 
a good approximation of the odd CSS. If squeezing is too large,
the CSS and squeezed single photon will become different.

\section{Non-deterministic CSS amplification process}

\begin{figure}
\centerline{\scalebox{0.48}{\includegraphics{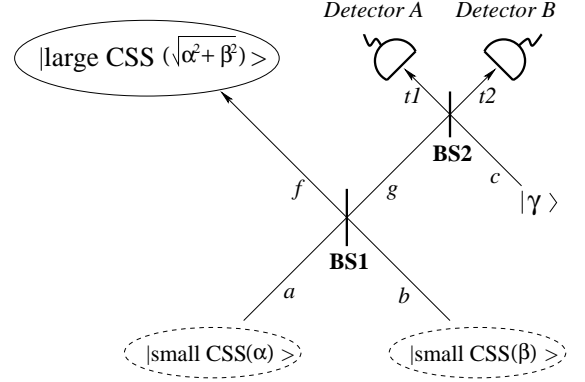}}}
\caption{
A schematic of the non-deterministic CSS-amplification process.
Two small CSSs at modes $a$ and $b$ are added to produce a larger
CSS at mode $f$ by a conditional measurement using detectors $A$ and $B$.  
See text for details.}
\label{fig-1}
\end{figure}

In this Section, we show that
 an arbitrarily large CSS can be produced
out of arbitrarily small CSSs using the simple experimental set-up 
depicted in Fig.~\ref{fig-1}.
Let us first illustrate this procedure with a simple example.
 Suppose that one has a collection of identical small odd CSSs with known 
amplitude $\alpha_i$.
Two of the small CSSs are selected and are incident onto a 
50:50 beam splitter, BS1, which acts on two coherent states 
$|\alpha\rangle$ and $|\beta\rangle$ as
\begin{equation}
|\alpha\rangle_a|\beta\rangle_b\stackrel{\rm BS1}{\longrightarrow}\
|\frac{\alpha}{\sqrt{2}}+\frac{\beta}{\sqrt{2}}\rangle_f
|-\frac{\alpha}{\sqrt{2}}+\frac{\beta}{\sqrt{2}}\rangle_g.
\end{equation}
The two small CSSs are then transformed at BS1 as
\begin{widetext}
\begin{eqnarray}
|{\rm CSS}_-(\alpha_i)\rangle_a|{\rm CSS}_-(\alpha_i)\rangle_b
&\stackrel{\rm BS1}{\longrightarrow}& 
|0\rangle_f\Big(|\sqrt{2}\alpha_i\rangle_g+|-\sqrt{2}\alpha_i\rangle_g\Big)
-\Big(|\sqrt{2}\alpha_i\rangle_f+|-\sqrt{2}\alpha_i\rangle_f\Big)
|0\rangle_g\nonumber\\
&\propto&|0\rangle_f|{\rm CSS}_+(\sqrt{2}\alpha)\rangle_g-|{\rm 
CSS}_+(\sqrt{2}\alpha)\rangle_f|0\rangle_g
\label{re-re-1}
\end{eqnarray}
\end{widetext}
where the normalization factors are omitted on the right hand side.
One can then say that if one could condition on detecting $|0\rangle_g$,
a larger CSS with amplitude $\sqrt{2}\alpha_i$ would be obtained at mode $f$.
However, the nonzero overlap between the vacuum and the even CSS
in Eq.~(\ref{re-re-1}) will make it impossible to perform unambiguous measurements. 
The error due to this overlap is not negligible because the initial amplitude
$\alpha_i$ is supposed to be a small value. Note that if the parity of the initial
CSSs are different, an unambiguous conditioning is possible using an ideal
photodetector. The two small CSSs of different parity are transformed at BS1 as
\begin{eqnarray}
\begin{aligned}
&|{\rm CSS}_-(\alpha_i)\rangle_a|{\rm CSS}_+(\alpha_i)\rangle_b\\
&~~~ \stackrel{\rm BS1}{\longrightarrow} 
|0\rangle_f|{\rm CSS}_-(\sqrt{2}\alpha)\rangle_g
+|{\rm CSS}_-(\sqrt{2}\alpha)\rangle_f|0\rangle_g.
\label{re-re-1.5}
\end{aligned}
\end{eqnarray}
where the normalization factor is omitted again on the right hand side.
Since the overlap between the vacuum and the odd CSS is zero, 
a larger odd CSS, $|{\rm CSS}_-(\sqrt{2}\alpha)\rangle_f$, can be conditionally produced
regardless of the value of $\alpha_i$
by detecting no photon at mode $g$.
Even in this case, however, the resulting states of conditional measurements will
be highly sensitive to detection inefficiency for small $\alpha_i$.
An additional step therefore is required to unambiguously discriminate between the vacuum and
coherent states $|\pm\sqrt{2}\alpha_i\rangle_g$ with inefficient detectors.
Another 50:50 beam splitter, BS2, mixes the field at mode $g$ and an auxiliary coherent state  
$|\sqrt{2}\alpha_i\rangle_c$ as
\begin{eqnarray}
|{\rm BS1}\rangle_{f,g}|\sqrt{2}\alpha_i\rangle_c
\stackrel{\rm BS2}{\longrightarrow} 
|0\rangle_f\Big(|2\alpha_i\rangle_{t1}|0\rangle_{t2}+|0\rangle_{t1}
|2\alpha_i\rangle_{t2}\Big)\nonumber
\\-\Big(|\sqrt{2}\alpha_i\rangle_f
+|-\sqrt{2}\alpha_i\rangle_f\Big)|\alpha_i\rangle_{t1}|-\alpha_i\rangle_{t2}~~~~
\label{re-re-2}
\end{eqnarray}
where $|{\rm BS1}\rangle_{f,g}$ represents the right hand side of Eq.~(\ref{re-re-1}) and
the normalization factor is omitted.
Finally, photodetectors $A$ and $B$ are set to detect photons 
at modes $t1$ and $t2$.
The remaining state at mode $f$ is selected
only when both the detectors detect any photon(s) at the same time.
In this case, it is obvious that the right hand side of Eq.~(\ref{re-re-2})
is reduced to a larger CSSs. 
If either of the detectors fails to click, the resulting state is discarded.

\begin{figure}
\centerline{\scalebox{0.43}{\includegraphics{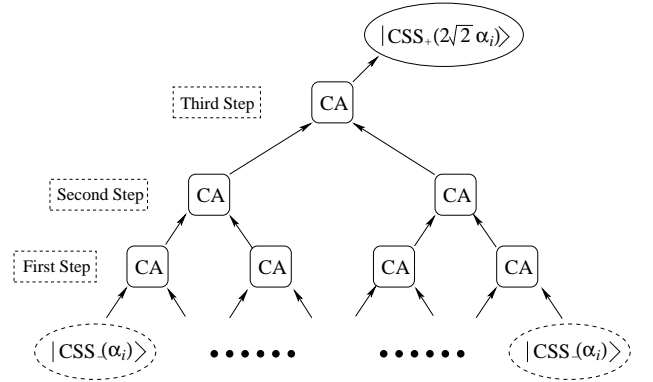}}}
\caption{
A schematic of successive applications of the 
CSS amplification processes.
CA represents the CSS amplification process depicted in Fig.~\ref{fig-1}.
As the first step, eight CSSs of initianl amplitude $\alpha_i$ are fed into the
four CA processes. If all the eight detectors in the four CA processes click,
the resulting states are selected for the second step, and so on.
In this example, 14 detectors and 14 beam splitters are required
with 7 auxiliary coherent states
to distill a CSS of amplitude $2\sqrt{2}\alpha_i$ from smaller CSSs of amplitude $\alpha_i$.}
\label{fig-new}
\end{figure}

This process can be 
successively applied until
a CSS of a sufficiently large amplitude is obtained.
Suppose that an even CSS with amplitude $\alpha>2$ is required
while the initial amplitude of small odd CSSs 
is $\alpha_i=1$.
One may consider an experimental setup
depicted in Fig.~\ref{fig-new} to obtain a CSS of the required amplitude.
Here we refer to the CSS amplification process depicted in Fig.~\ref{fig-1} as ``CA''.
Firstly, four pairs of odd CSSs (i.e. eight odd CSSs) with amplitude 
$\alpha_i=1$ should be fed
into four CA processes simultaneously as shown in Fig.~\ref{fig-new}.
If the first step in Fig.~\ref{fig-new} is successful, i.e.,
all the eight detectors in the first four CA processes click,
two pairs of even CSSs with amplitude $\sqrt{2}\alpha_i$
will be generated out of the four pairs of smaller CSSs
of amplitude $\alpha_i$
fed into the first four CA processes.
In the second step,
two CA processes are performed 
with the two pairs of even CSSs generated from the first step.
Note that the auxiliary coherent states 
for the second step should be $|2\alpha_i\rangle$'s.
Through this second stage,
one pair of even CSSs of amplitude $2\alpha_i$ 
can be gained from the two pairs of even CSSs with amplitude $\sqrt{2}\alpha_i$.
Finally, an even CSS with amplitude $2\sqrt{2}\alpha_i(\approx2.83)$ 
can be generated by the third step
which is only a single CA process with an appropriate auxiliary state. 
An CSS of an arbitrarily larger amplitude can be produced 
by increasing the number of the steps from any smaller CSSs.
Of course, the success probability will rapidly drop down
and the required resources will exponentially 
increase as the number of steps increases
unless quantum optical memory is available.

\begin{figure}
\centerline{\scalebox{0.63}{\includegraphics{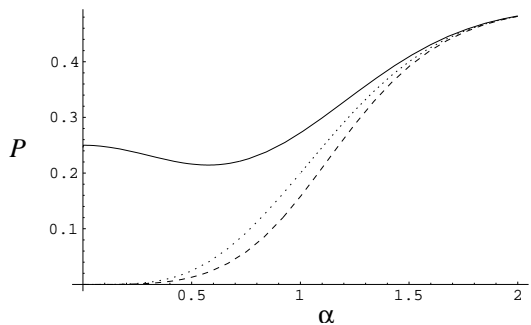}}}
\caption{The success probabilities of the CSS-amplifying process in
Fig.~\ref{fig-1} for the input fields of two identical odd CSSs (solid line),
two identical even CSSs (dashed line), and even and odd CSSs (dotted line).}
\label{fig:prob}
\end{figure}

The CA process described above can be generalized for
arbitrarily small CSSs with 
known amplitudes as already shown in  Fig.~\ref{fig-1}.
Suppose two small CSSs, 
$|{\rm CSS}_\varphi(\alpha)\rangle$ and $|{\rm CSS}_\phi(\beta)\rangle$,
with amplitudes $\alpha$ and $\beta$. 
The reflectivity $r$ and transmitivity $t$ of BS1
 are set to $r=\beta/\sqrt{\alpha^2+\beta^2}$ and 
$t=\alpha/\sqrt{\alpha^2+\beta^2}$, where 
the action of the beam splitter is represented by
\begin{equation}
{\hat B}_{a,b}(r,t)|\alpha\rangle_a|\beta\rangle_b
|t\alpha+r\beta\rangle_f|-r\alpha+t\beta\rangle_g.
\end{equation}
The other beam splitter BS2 is a 50:50 beam splitter ($r=t=1/\sqrt{2}$)
regardless of the conditions and
the amplitude $\gamma$ of the auxiliary coherent field is 
determined as 
\begin{equation}
\gamma=2\alpha\beta/\sqrt{\alpha^2+\beta^2}.
\end{equation}
The two beam splitters BS1 and BS2 then transform the two small CSSs with the
auxiliary state as
\begin{widetext}
\begin{eqnarray}
\begin{aligned}
&{\hat B}_{g,c}\Big(\frac{1}{\sqrt{2}},\frac{1}{\sqrt{2}}\Big){\hat B}_{a,b}
\Big(\frac{\beta}{\sqrt{\alpha^2+\beta^2}},\frac{\alpha}
{\sqrt{\alpha^2+\beta^2}}\Big)|{\rm CSS}_\varphi(\alpha)\rangle_a|{\rm CSS}_\phi
(\beta)\rangle_b|\gamma\rangle_c=\\
&\Bigg\{\Big(|{\cal A}\rangle+e^{i(\varphi+\phi)}|-{\cal A}\rangle\Big)
|\frac{\gamma}{\sqrt{2}}\rangle
|\frac{\gamma}{\sqrt{2}}\rangle
+e^{i\phi}|\frac{\alpha^2-\beta^2}{\cal A}\rangle
|0\rangle|\sqrt{2}\gamma\rangle+e^{i\varphi}|-\frac{\alpha^2-\beta^2}
{\cal A}\rangle|\sqrt{2}\gamma\rangle|0\rangle\Bigg\}_{f,t1,t2}
\equiv|\Phi\rangle,
\label{bs12}
\end{aligned}
\end{eqnarray}
\end{widetext}
where ${\cal A}=\sqrt{\alpha^2+\beta^2}$.
Here, the measurement operator $\hat{P}_{t1,t2}$ can be represented
 as 
\begin{equation}
\hat{P}_{t1,t2}=(\openone-|0\rangle\langle0|)_{t1}\otimes
(\openone-|0\rangle\langle0|)_{t2}. 
\end{equation}
It is then obvious from Eq.~(\ref{bs12}) 
that the resulting state for mode $f$ 
by the ``click-click'' event at $t1$ and $t2$
becomes 
$|{\rm CSS}_{\varphi+\phi}({\cal A})\rangle\propto|{\cal A}\rangle
+e^{i(\varphi+\phi)}|-{\cal A}\rangle$,
whose coherent amplitude ${\cal A}=\sqrt{\alpha^2+\beta^2}$
is larger than both $\alpha$ and $\beta$.
The relative phase of the resulting CSS
is the sum of the relative phases of the input CSSs.
The success probability $P_{\varphi,\phi}(\alpha,\beta)$ for a single iteration of
the process above is simply calculated as
\begin{eqnarray}
P_{\varphi,\phi}(\alpha,\beta)&=&
\langle\Phi| \hat{P}_{t1,t2} |\Phi\rangle\nonumber\\
&=&\frac{
(1-e^{-\frac{2\alpha^2\beta^2}{\alpha^2+\beta^2}})^2
[1+\cos(\varphi+\phi)e^{-2(\alpha^2+\beta^2)}]}
{2(1+\cos\varphi e^{-2\alpha^2})(1+\cos\phi e^{-2\beta^2})},\nonumber
\end{eqnarray}
which is plotted for a number of different combinations 
in Fig.~\ref{fig:prob}.
The success probabilities depend on the type of CSSs (odd or even) used
and it approaches 1/2 as the amplitudes of the initial CSSs becomes large.
It is interesting to note that the probability $P_{\pi,\pi}(\alpha,\alpha)$ for two
identical odd CSS inputs is always larger than $\sim0.214$ regardless of the
value of $\alpha$ as shown in Fig.~\ref{fig:prob}. This is due to the fact that each
odd CSS contains at least one photon no matter how small its amplitude is.
Multiple iterations will rapidly reduce the success probability. 
For example, if one needs to distill a CSS of $\alpha=2$ out of 4 CSSs of
$\alpha=1$ ($\alpha=1/\sqrt{2}$), the success probability will be
$\sim 0.027$ ($4\times10^{-4}$).
If a CSS of $\alpha=2$ is desired out of 16 odd CSSs of
$\alpha=1/2$, the success probability will be only $2\times10^{-13}$.
However, if quantum memory is available one can temporarily hold the 
output state upon success waiting for the remainder of the trials to 
give a successful result. This avoids the exponential scaling of the 
overall probability of success and for the $\alpha = 2$ case just 
considered the average number of steps is 1731.
The inefficiency of photodetectors
will also decrease the success probability while it does not
affect the quality of the obtained CSSs.

It is worth noting that not only an arbitrary large even CSS but
also an arbitrarily large odd CSS can be obtained out of small odd CSSs.
If a larger odd CSS needs to be produced, a larger even CSS 
obtained from a collection of initial odd CSSs and a single initial
odd CSS can be fed into a CA process 
so that a larger odd CSS can be obtained. If the even CSS of amplitude $2\sqrt{2}$
obtained and the initial odd cat of amplitude $1$
are used as the two input states in Fig.~\ref{fig-1},
an odd CSS of amplitude $3$ will then be obtained.

\begin{figure}
\centerline{\scalebox{0.7}{\includegraphics{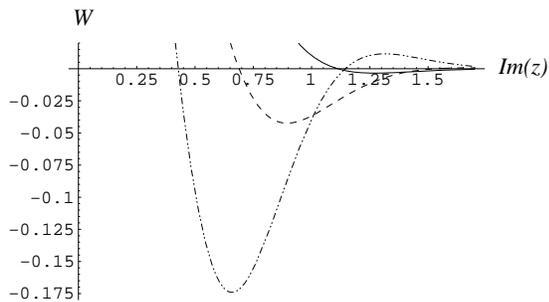}}}
\caption{A section of the Wigner function for an initial even CSS of amplitude 1/2 (solid 
line),
the resulting CSS after the first iteration (dashed line), and the resulting state
after the second iteration (dotted line). The maximum negative
values are shown in the figure. One can see a radical increase of negativity of
the Wigner function.}
\label{add1}
\end{figure}

We have pointed out that a small CSS approaches a single photon while
a larger CSS is a superposition of macroscopically distinguishable
states which can be considered a
realization of Schrodinger's paradox.
This means that in principle, our scheme 
distills a superposition of macroscopically distinguishable states
from microscopic quantum states.
It also increases
non-classical features of quantum states.
The negativity of Wigner functions is an indicator of non-classical
features of a quantum state. 
Since an even CSS approaches the vacuum state as its amplitude gets smaller
and the Wigner function of the vacuum state is positive-definite,
the maximum negative value of the Wigner function is small for a small even CSS.
In this regime of a small amplitude, the Wigner function of an even CSS
looks almost like a Gaussian state.
Fig.~\ref{add1} shows how the maximum
negative value of the Wigner function of an even CSS
increases as our amplification process is iterated.

\section{Amplifying squeezed single photons for larger coherent state superpositions }

In our earlier discussions, it was shown that the fidelity between
a squeezed single photon and an ideal small cat
is extremely high.  
Therefore, it can be conjectured that 
a larger CSS distilled from squeezed single photons by our scheme will 
also be very close to an ideal CSS.
 In what follows, we will show that this conjecture is right for $\alpha\leq2.5$
by analytical and numerical approximations.
We first choose the initial coherent amplitude as $\alpha_i=1/\sqrt{2}$.
The fidelity of the initial CSS, which is the squeezed single photon,
is then $F=0.99978$ for the appropriate squeezing parameter $r=0.163725$.

The squeezed single photon can be represented in terms of an ideal CSS
and error components as
$S(r)|1\rangle 
\propto|CSS_-(\frac{1}{\sqrt{2}})\rangle+\delta^{(3)}
|3\rangle+\delta^{(5)}|5\rangle
+\delta^{(7)}|7\rangle+\cdot\cdot\cdot$, 
where the error terms are
\begin{equation}
\delta^{(2k+1)}=\frac{e^{1/4}\Big[0.162278^k(2k+1)!-k!\Big]}{2^k k!\sqrt{(2k+1)!(e-1)}}.
\end{equation}
It can be simply checked that $\delta^{(5)}=0.0129669$ 
is the dominant error term and $\delta^{(k)}$ exponentially 
decreases for $k>5$. 
The  state only with $\delta^{(5)}$, 
 $N(|CSS_-(\frac{1}{\sqrt{2}})\rangle+\delta^{(5)}|5\rangle)$,
  where $N$ is the normalization factor, will give a fidelity $F=0.99983$ for 
  the odd CSS $|CSS_-(\frac{1}{\sqrt{2}})\rangle$.
  In other words,
the state only
with the dominant error term can be a good approximation of the squeezed single photon for a 
weak squeezing. We therefore use 
\begin{equation}
S(r)|1\rangle\approx|\Psi_i
\rangle=N_i(|CSS_-(1/\sqrt{2})\rangle+\delta_i|5\rangle)
\end{equation}
as the initial input state, where $r=0.163725$, $\delta_i=0.0147$ and $N_i$ is a normalization 
factor. The initial fidelity $F_i$ between
$|\Psi_i\rangle$ and the ideal odd CSS is made $F_i=0.99978$, which is exactly same as the 
case of a squeezed single photon.
The resulting state is obtained as
\begin{eqnarray}
&&|\Psi_{(1)}\rangle_{f,t1,t2}=\hat{P}_{t1,t2}\hat{B}^{1:1}_{g,c}
 \hat{B}^{1:1}_{a,b}|\Psi_i\rangle_{a}|\Psi_i\rangle_{b}|\sqrt{2}
\alpha_i\rangle_c,~~~~~\\
&&{\rho_{(1)}}_f={\rm Tr}_{t1,t2}\big[|\Psi_{(1)}\rangle\langle\Psi_{(1)}|\big],~~~~~
\end{eqnarray}
where ${\rm Tr}_{t1,t2}$ denotes a partial trace of modes $t1$ and $t2$, 
a subscript $(n)$ indicates the number of the 
iterative steps made to amplify
the CSS, and $\hat{B}^{1:1}=\hat{B}(\frac{1}{\sqrt{2}},\frac{1}{\sqrt{2}})$.

When each of the detectors detects only one photon, it is
straightforward to calculate the resulting state
\begin{equation}
_{t1}\langle 1|_{t2}\langle1|\Psi_{(1)}\rangle
\propto|CSS_+(1)\rangle-1.11 \delta_i\Big(0.828|4\rangle-0.561|6\rangle\Big),
\label{eq:11}
\end{equation}
where 
 error terms smaller than 1/3 of the dominant error term have been discarded as
they  exponentially decay. 
Considering the normalization, the fidelity of the state (\ref{eq:11}) 
is calculated to be 0.99974.
Note that about 60\% of all the 
successful simultaneous clicks at detectors $A$ and $B$ 
correspond to this case, where the probability can be calculated
by $P^{n,m}_{(n)}=\langle n |\langle m|{\rho_{(n)}}_f|n\rangle|m\rangle$.
About 30\% of the successful clicks correspond to the cases that 
detector $A$ detects two photons while detector $B$ detects one photon
or that $A$ detects one while $B$ detects two photons.
 We can make a same approximation for these cases and
the fidelity is 0.99975.
On the other hand, the highest overlap with a CSS of $\alpha=1$ that can be
obtained by simply squeezing a single photon is $F=0.99711$, thus a clear improvement has been 
obtained.

\begin{figure}
\centerline{\scalebox{0.49}{\includegraphics{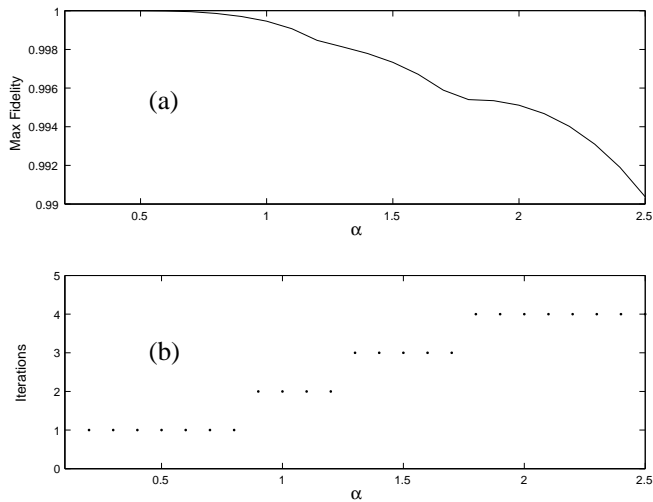}}}
\caption{(a) The maximum fidelity obtained in our scheme vs the coherent
 amplitude. (b) The number of iterations which gives the maximum fidelity vs
 the coherent amplitude. The improvements of the fidelity
 are remarkable when compared with Fig.~\ref{fid-nopa}.}
\label{something}
\end{figure}

In order to calculate multiple iterations we need to use numerical techniques.
 We are using coherent states of
 some bounded coherent amplitude and superpositions there of. Provided the
 coherent amplitudes are not small, the most significant contributions
 to these states are Fock states of low number.  For computations here 
the lowest thirty Fock states were used.  This provides a very good
 approximation for coherent states with $\alpha \leq 2.5$.  
All 29 possible ``click'' events are included 
for all detectors.   

If one wished to create a CSS with a particular $\alpha$ with $n$ 
CSS-amplification steps, then initial CSSs with 
$\alpha_i = \alpha/\sqrt{2}^n$ are required.
 As the number of steps increases the required $\alpha_i$ decreases.  
 When generating a larger CSS 
out of the squeezed single photon states 
  the fidelity maximizes for
 a particular number of iterations.  Fig.~\ref{something} shows the maximum
 possible fidelity using this process in (a) and the number of steps
 in (b) against the desired $\alpha$ in the CSSs.
For example, four iterations starting from the initial amplitude $\alpha_i=1/2$ is required to
gain the maximum fidelity $F=0.995$ for $\alpha=2$.
It is evident from Fig.~4 that high fidelity, $F>0.99$, can be obtained up to $\alpha= 2.5$.
The error rate for discrimination between coherent states with $\alpha=\pm 2.5$
via a classical measurement (homodyne detection) is only $3\times10^{-7}$.

\section{Purification effects of the CSS amplification process}

The single photons required for our scheme could be generated conditionally
from a down-converter \cite{Lvovsky}. This is a $\chi^{(2)}$ process
(like squeezing) and does not require photon number resolving detection.
It should be noted that current technology does not produce pure single photon states;
the single photon is always in a mixture with the vacuum as 
\begin{equation}
p\ket{0}\bra{0}+ (1-p)\ket{1}\bra{1},\label{mixedstate}
\end{equation}
where $p$ is the inefficiency of the photon production.
Hence the squeezed single photon state will also be a mixture with a squeezed vacuum as
\begin{equation}
p\hat S(r)\ket{0}\bra{0}\hat S^\dagger(r)+ 
(1-p)\hat S(r)\ket{1}\bra{1}S^\dagger(r).
\label{mis1}
\end{equation}
However, an interesting aspect of our scheme is that
it may be somewhat resilient to the photon production inefficiency because
 its first iteration purifies the mixed CSSs while amplifying them.
The initial input states for the CSS amplification process 
from the imperfect single photon source are
\begin{widetext}
\begin{eqnarray}
&&\rho_{a,b,c}
=\Big[(1-p)^2\ket{S_1}\bra{S_1}\otimes\ket{S_1} \bra{S_1}
+p^2\ket{S_0} \bra{S_0}\otimes\ket{S_0}\bra{S_0}
\nonumber\\
&&~~~~~~~~~~~~
+ p(1-p) \Big(\ket{S_0} \bra{S_0}\otimes \ket{S_1} \bra{S_1}
+\ket{S_1} \bra{S_1}\otimes \ket{S_0} \bra{S_0}\Big)\Big]_{a,b}\otimes\big(
|\gamma\rangle\langle\gamma|\big)_c
\label{mis}
\end{eqnarray} 
\end{widetext}
where $|S_0\rangle =\hat{S}(r) \ket{0}$ and $|S_1\rangle =\hat{S}(r) \ket{1}$.
Here, the terms with $p^2$ and $p(1-p)$ are undesired error terms where either  (or both) of
the single photons  is missing.
Note that the initial amplitude is required to be small to produce a larger CSS with high 
fidelity.
Provided such a small amplitude,
input states incident onto the beam splitters in our experimental setup
contain approximately only two (or slightly more than two) photons.
In such cases the probability of simultaneous clicks at detectors $A$ and $B$
in Fig.~\ref{fig-1} will significantly decrease when any of the single photons is missing.
In other words, the undesired cases will rarely be selected 
for the next iteration of the amplification process.
We have obtained numerical results 
for the initial amplitude $\alpha_i=1/2$ as follows by the methods that we have already
explained. If $p=0.4$, the fidelity of the initial CSS, which is a mixture with a squeezed 
vacuum,
is $F=0.60$ but it will become $F=0.89$ by the first iteration.
Thus a larger CSS of significantly high fidelity is produced.
If $p=0.25$ $(p=0.05)$, the fidelity of the initial CSS is $F=0.750$ $(F=0.950)$ but
 becomes $F=0.941$ $(F=0.990)$ by the first iteration.

 \begin{figure}
\centerline{\scalebox{0.5}{\includegraphics{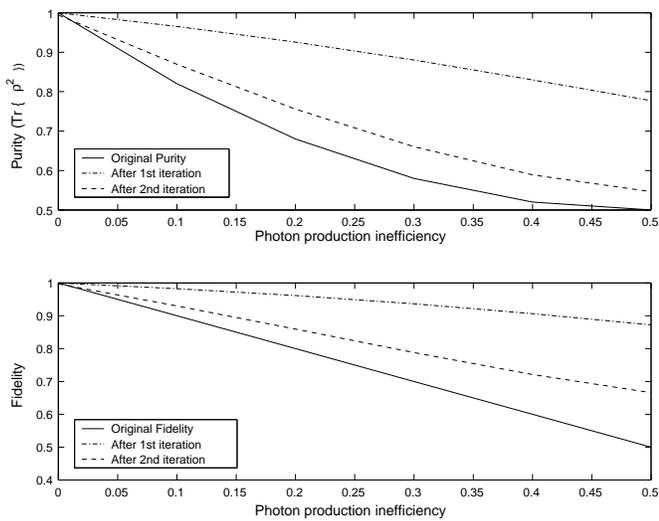}}}
\caption{(a) Purity and (b) fidelity of the input state (solid lines),
the input output state after the first iteration (dot-dashed lines),
 and of the output after the second iteration (dashed line)
 as functions of 
the photon production inefficiency $p$.
The input states are mixed squeezed single photons 
intended to be approximate odd CSSs of amplitude $\alpha = 1/2$.
The final output states should be therefore approximate CSSs of $\alpha = 1$. 
}
\label{doublecatpurity1}
\end{figure}

The purification by the first iteration is directly evident by the probability argument,
but what remains to be shown is if this effect is strong enough for purification to still
be achieved for multiple iterations. 
For a double iteration, four input CSSs of $\alpha_i=1/2$ would be required to
obtain an output CSS of $\alpha=1$.
The numerical results of the second iteration presented in Fig.~\ref{doublecatpurity1}
show that the improvements of fidelity
obtained by the first step
can remain for further iterations.
In Fig.~\ref{doublecatpurity1}(a), 
the purity of the input state (solid line) and the purity of
the output state after the first (dot-dashed line)
and second (dashed line) iterations  
have been plotted as functions of 
the photon production inefficiency $p$.
Note that purity is defined here as ${\rm Tr}\{\hat{\rho}^2\}$.
Fig.~\ref{doublecatpurity1}(b) 
shows the fidelity of the input state when compared with the
ideal CSS of $\alpha=1/2$ (solid line), the fidelity of the output state
after the first iteration compared with the CSS of $\alpha=1/\sqrt{2}$
(dot-dashed line), 
the fidelity of the output state (dashed line) compared with the
ideal CSS of $\alpha=1$. 
When $p=0$ the purity and fidelity are high as the input states themselves are pure and the 
output fidelity is expected to be high.
The fidelity has decreased compared with the result after the first iteration but it is still
higher than the initial fidelity. 
For example, the fidelity will change from 0.60 to 0.89 for the first iteration and finally 
to 0.72 for the second iteration when $p=0.4$.  
For the range of probabilities shown here ($p \in [0,0.5]$) there is always an
improvement in purity and fidelity.

There exists an alternate way of achieving the same output state using the same input CSSs but 
using a different arrangement of amplification procedures.  Two $\alpha = 1/2$ odd input CSSs 
could be amplified to generate one even $\alpha = \sqrt{1/2}$ CSS.  Then this state could be 
combined with another $\alpha = 1/2$ odd CSS to create a $\alpha = \sqrt{3/4}$ odd CSS.  Then 
finally this state could be combined with another $\alpha = 1/2$ odd CSS to generate the 
$\alpha = 1$ even CSS output.
One might expect that the presence of the $\alpha = \sqrt{3/4}$ odd CSS could make
some differences in purity and fidelity by this method.
However, the plots for the final output state which we have obtained
by the same numerical technique 
are identical in nature to
those of Fig.~\ref{doublecatpurity1}. 
So the purifying effects of this procedure
do not seem to depend on the way in which the output state is generated.

\section{Remarks}

We have studied a simple all-optical scheme to generate
a linear superposition of macroscopically distinguishable
coherent states in a propagating optical field \cite{Lund04}.
In stark contrast to all previous schemes, this scheme 
requires neither $\chi^{(3)}$ nonlinearity nor efficient photon detection
to generate
a superposition of macroscopically distinguishable states.
Furthermore,
it exhibits some resilience to photon production inefficiency
because it purifies initial mixed states. We 
have found that these purification effects can last for multiple iterations.  
The non-deterministic CSS amplification scheme
has been proved to boost non-classicality of quantum states:
even very small amount of negativity can be drastically
increased by this process.
This scheme non-deterministically generates CSSs with amplitude $\alpha>2$.
However, it should be noted that 
a non-deterministic CSS source is useful enough for quantum
information processing \cite{Ralph,Ralph2}. 

In the CSS amplification process, 
the zero amplitude coherent states that occur in the detection modes
in Eq.~(\ref{re-re-2})
 may be slightly different from zero
because of imperfect mode matching 
 at beam splitters.
This will lead to a small probability of accepting the wrong state.
Good mode matching is a requirement in any linear optical network where one wishes
to measure manifestly quantum mechanical effects.
Highly efficient mode matching of a single photon
from parametric down conversion and a weak coherent state from an attenuated laser beam 
at a beam splitter has been experimentally demonstrated
using optical fibers \cite{Pitt}. Such techniques could be employed for
the implementation of our scheme.

The dark count rate of photodetectors will affect the fidelity of the CSSs.
Currently, highly efficient detectors have relatively high dark count
rates while less efficient detectors have very low dark count rates
\cite{Takeuchi}. We emphasize again that our scheme does not require
highly efficient detectors because the inefficiency of the detectors
does not affect the quality of CSSs even though it decreases the success
probability. Silicon avalanche photodiodes operating at the visible wavelength have
relatively high efficiency and a small dark count rate, which is
preferred in our proposal.

Once free propagaing CSSs states are generated, they can be 
detected by homodyne measurements, which can be highly efficient 
in quantum optics experiments. Interference fringes will appear
as a signature of the CSSs in the statistics of the photocurrent at the detectors.

Finally, we note that there is an alternative method to obtain a squeezed single
photon even without a single photon source. It is perhaps not surprising
that a squeezed single photon can be obtained by adding a photon to
a squeezed vacuum as
$\hat a^\dagger S(r)|0\rangle=\sinh r S(r)|1\rangle$.
However, an interesting observation is that a squeezed single photon can also be
obtained by subtracting a photon from a squeezed vacuum.
This can be shown by applying the annihilation operator to a squeezed
single photon: 
\begin{equation}
\hat a S(r)|0\rangle=\cosh r S(r)|1\rangle.
\label{sspm}
\end{equation}
It was already pointed out that a photon-subtracted or photon-added
squeezed vacuum state
is similar to a CSS \cite{pst}.
Recently, a free-propagating non-Gaussian optical state
which is close to a squeezed single photon in Eq.~(\ref{sspm}) 
has been experimentally deomonstrated
by Wenger {\it et al.} \cite{Wenger}.
In their experiment, the single photon subtraction was approximated
by a beam splitter of low reflectivity and a single photon detector.
Such an experiment could be immediately linked to our suggestion to
experimentally generate a larger CSS. One can then generate 
a CSS of $\alpha>2$ using our scheme without a single photon source.

We would like to thank M.S. Kim for useful comments.
This work was supported by the Australian Research Council and the
University of Queensland Excellence Foundation.


\begin{thebibliography} {}



\bibitem{Schr} 
E. Schr$\rm\ddot{o}$dinger, {\it Naturwissenschaften.} {\bf 23},  807-812; 823-828; 844-849 
(1935).

\bibitem{Leggett} A.J. Leggett and A. Garg, \prl {\bf 54}, 857 (1985). 

\bibitem{Reid} 
M.D. Reid, Quantum Semiclass. Opt. {\bf 9}, 489 (1997);
M.D. Reid, \prl {\bf 84}, 2765 (2000);
 M.D. Reid, \pra {\bf 62}, 022110 (2000);
 M.D. Reid, quant-ph/0101052.

\bibitem{Enk} S.J. van Enk and O. Hirota,  Phys. Rev. A {\bf 64}, 022313 (2001).

\bibitem{JKL01} H. Jeong, M.S. Kim, and J. Lee, Phys. Rev. A {\bf 64}, 052308 (2001).

\bibitem{Wang} X. Wang, Phys. Rev. A {\bf 64}, 022302 (2001).

\bibitem{BaAnTeleport} Nguyen Ba An, Phys. Rev. A {\bf 68}, 022321 (2003).

\bibitem{BaAnW} Nguyen Ba An, Phys. Rev. A {\bf 69}, 022315 (2004).

\bibitem{JK} H. Jeong and M.S. Kim, Phys. Rev. A {\bf 65}, 042305 (2002).

\bibitem{Ralph} T.C. Ralph, W.J. Munro, and G.J. Milburn,
Proceedings of SPIE {\bf 4917}, 1 (2002);  quant-ph/0110115.

\bibitem{Ralph2} T.C. Ralph, A. Gilchrist, G.J. Milburn,
W.J. Munro, and S. Glancy, Phys. Rev. A 68, 042319 (2003).

\bibitem{JKpuri} H. Jeong and M.S. Kim,  Quantum Information and Computation
 {\bf 2}, 208 (2002); J. Clausen, L. Kn{\" o}ll, and D.-G. Welsch,
Phys. Rev. A {\bf 66}, 062303 (2002).

\bibitem{Glancy}  P.T. Cochrane, G.J. Milburn, and W.J. Munro,
 Phys. Rev. A {\bf 59}, 2631 (1999); S. Glancy, H.M. Vasconcelos,
 and T.C. Ralph, Phys. Rev. A {\bf 70}, 022317 (2004). 

\bibitem{Yurke} B. Yurke and D. Stoler, \prl {\bf 57}, 13 (1986).

\bibitem{Boyd} R.W. Boyd, J. Mod. Opt {\bf 46}, 367 (1999).

\bibitem{Song} S. Song, C.M. Caves, and B. Yurke, Phys. Rev. A {\bf
    41}, R5261 (1990).
    
\bibitem{Dakna} M. Dakna, T. Anhut, T. Opatrn{\' y}, L. Kn\"oll,
 and D.-G. Welsch, Phys. Rev. A {\bf  55}, 3184 (1997).
  
\bibitem{Dakna2} M. Dakna, J. Clausen, L. Kn\"oll and D. -G. Welsch,
  Phys. Rev. A {\bf  59}, 1658 (1999);
  J. Clausen, M. Dakna, L. Kn\"oll and D.-G. Welsch, Optics 
Communications {\bf 179}, 189 (2000).
  
\bibitem{Dakna3} J. Clausen, M. Dakna, L. Kn\"oll and D. -G. Welsh,
 Acta Physica Slovaca {\bf 49}, 96 (1999) quant-ph/9905085.

\bibitem{Montina} A. Montina and F.T. Arecchi, \pra {\bf 58}, 3472 (1998).

\bibitem{Tu} Q.A. Turchette, C.J. Hood, W. Lange, H. Mabuchi, and
  H.J. Kimble, Phys. Rev. Lett. {\bf 75}, 4710 (1995).

\bibitem{MB} M. Brune, E. Hagley, J. Dreyer, X. Ma{\^ i}tre, A. Maali,
  C. Wunderlich, J.M. Raimond, and S. Haroche, Phys. Rev. Lett. {\bf 77},
  4887 (1996).
  
\bibitem{Mon} C. Monroe, D.M. Meekhof, B.E. King, and D.J. Wineland,
  Science {\bf 272}, 1131 (1996).

\bibitem{Lund04} A.P. Lund, H. Jeong, T.C. Ralph, and M.S. Kim,
Phys. Rev. A {\bf 70}, 020101(R) (2004).
  
\bibitem{pst} D.-G. Welsch, M. Dakna, L. Kn\"oll and T. Opatrny,
5th International Conference on Squeezed States and Uncertainty 
Relations (Balatonfured, Hungary, 1997) Proceedings: NASA/CP-1998-206855, 
609 (1998) and references therein.

\bibitem{Wenger} J. Wenger, R. Tualle-Brouri, and P. Grangier, Phys. Rev. Lett {\bf 92},
153601 (2004).
	
\bibitem{Schr2}   E. Schr\"odinger, Naturwissenschaften 14, 664 (1926).
  
\bibitem{BRbook} S.M. Barnett and P.M. Radmore, 
{\it Methods in Theoretical Quantum Optics}, Oxford University Press, New York (1997). 
(1926).
  
\bibitem{Barnett}  S.M. Barnett, C.R. Gilson, and M. Sasaki,
quant-ph/0209138.  
  
\bibitem{coherentQKD} S. Iblisdir, G. Van Assche and N. J. Cerf, quant-ph/0312018.
  
\bibitem{coherent} L.M. Johansen, quant-ph/0309025.

\bibitem{Martini} F. De Martini, Phys. Rev. Lett {\bf 81}, 2842 (1998);
F. De Martini, M. Fortunato, P. Tombesi, and D. Vitali, Phys. Rev. A
{\bf 60}, 1636 (1999).

\bibitem{LaPorta} A. La Porta, R. E. Slusher and B. Yurke, Phys. Rev. Lett. {\bf 62}, 28 
(1989).

\bibitem{APLThesis}  A.P. Lund, Honours Thesis,
Department of Physics, University of Queensland, Brisbane (2004).  

\bibitem{Branczyk} A.M. Bra\'nczyk, T.J. Osborne, A. Gilchrist,
and T.C. Ralph, Phys. Rev. A. {\bf 68}, 043821 (2003).
  
  
\bibitem{memory}  B. Julsgaard, J. Sherson, J.I. Cirac, J. Fiurasek, E.S. Polzik,
quant-ph/0410072 and references therein.
  
\bibitem{Lvovsky} A.I. Lvovsky, H. Hansen, T. Aichele, O. Benson, J. Mlynek,
and S. Schiller, Phys. Rev. Lett. {\bf 87}, 050402 (2001).

\bibitem{Takeuchi} S. Takeuchi, J. Kim, Y. Yamamoto, and H.H. Hogue (1999), 
 Appl. Phys. Lett. {\bf 74}, 1063.
 
\bibitem{Pitt} T.B. Pittman and J.D. Franson, 
\prl {\bf 90}, 240401 (2003).

\end{thebibliography}
\end{document}